\documentclass[useAMS,twocolumn,natbib]{svjour3}
\pdfoutput=1
\usepackage{amsmath}
\usepackage{amssymb}
\usepackage{graphicx}
\usepackage{tabularx}
\usepackage{booktabs}
\usepackage{multirow}
\usepackage{array}
\usepackage{caption}
\usepackage{siunitx}
\usepackage{float}
\usepackage{placeins}
\usepackage{color}
\usepackage[lined,linesnumbered,commentsnumbered]{algorithm2e}
\usepackage{hyperref}
\usepackage{soul}
\definecolor{red}{rgb}{0.9,0,0}
\definecolor{green}{rgb}{0,0.7,0}

\newdimen\myvskip
\newcolumntype{L}[1]{>{\raggedright\let\newline\\\arraybackslash\hspace{0pt}}m{#1}}
\newcolumntype{C}[1]{>{\centering\let\newline\\\arraybackslash\hspace{0pt}}m{#1}}
\newcolumntype{R}[1]{>{\raggedleft\let\newline\\\arraybackslash\hspace{0pt}}m{#1}}
\DeclareSIUnit\Msun{\ensuremath{\mathrm M_{\odot}}}
\DeclareSIUnit\cm{\ensuremath{\mathrm{cm}}}
\DeclareSIUnit\year{\ensuremath{\mathrm{yr}}}
\DeclareSIUnit\parsec{\ensuremath{\mathrm{pc}}}

\begin{document}

\title{PHEW: a parallel segmentation algorithm for three-dimensional AMR datasets}
\subtitle{ \it \\Application to structure detection in self-gravitating flows}
\author{A. Bleuler \and R. Teyssier \and S. Carassou \and D. Martizzi}
\institute{A. Bleuler \and R. Teyssier \and S. Carassou \and D. Martizzi \at
Institute for Computational Science, University of Zurich, CH-8057 Zurich, Switzerland 
\and 
S. Carassou \at
Institut d'Astrophysique de Paris, 98bis boulevard Arago, 75014 Paris, France
\and 
D. Martizzi \at
Department of Astronomy, University of California, Berkeley, CA 94720-3411, USA
\\
E-mail: ableuler@physik.uzh.ch }

%\date{Accepted Year Month Day. Received Year Month Day; in original form Year Month Day}
%\pagerange{\pageref{firstpage}--\pageref{lastpage}} \pubyear{2014}
\label{firstpage}
\maketitle

\begin{abstract}
We introduce \textsc{phew} (\textbf{P}arallel \textbf{H}i\textbf{E}rarchical \textbf{W}atershed), a new segmentation algorithm to detect structures in astrophysical fluid simulations, and its implementation into the adaptive mesh refinement (AMR) code \textsc{ramses}. \textsc{phew} works on the density field defined on the adaptive mesh, and can thus be used on the gas density or the dark matter density after a projection of the particles onto the grid. The algorithm is based on a "watershed" segmentation of the computational volume into dense regions, followed by a merging of the segmented patches based on the saddle point topology of the density field. \textsc{phew} is capable of automatically detecting connected regions above the adopted density threshold, as well as the entire set of substructures within. Our algorithm is fully parallel and uses the MPI library. We describe in great detail the parallel algorithm and perform a scaling experiment which proves the capability of \textsc{phew} to run efficiently on massively parallel systems. Future work will add a particle unbinding procedure and the calculation of halo properties onto our segmentation algorithm, thus expanding the scope of \textsc{phew} to genuine halo finding.
\end{abstract}

%\begin{keywords}
%methods: numerical
%\end{keywords}

%%%%%%%%%%%%%%%%%%%%%%%%
%%%%%%%%%%%%%%%%%%%%%%%%
%%%%%%%%%%%%%%%%%%%%%%%%
%INTRODUCTIONf 
%%%%%%%%%%%%%%%%%%%%%%%%
%%%%%%%%%%%%%%%%%%%%%%%%
%%%%%%%%%%%%%%%%%%%%%%%%
\section{Introduction}
Over the last decades, computer simulations have become an indispensable tool for studying the formation of structure on all scales in our universe. The common feature of those simulations is the clustering of matter due to self gravity. This clustering is of fractal nature in the sense that - as long as gravity is the dominant force - aggregations of matter turn out to have internal substructures, which are themselves gravitational bound, and may even contain sub- substructures. A crucial tasks in the analysis of simulations is therefore the identification of overdense regions and, ideally, their entire hierarchy of substructure.

First algorithms to perform this task have been invented in the very early days of computer simulations in Astronomy and Astrophysics. A halo finder based on spherical overdensities (SO) was described already four decades ago by \cite{Press1974} who used it to find structure in their simulation of 1000 particles. Subsequently, the SO method has become one of the standard methods for halo finding. It consists in growing spherical regions around density peaks and assigning particles inside the spheres to the respective peak based on physical arguments. The also very popular friends-of-friends (FOF) method was introduced to halo finding by \cite{davis1985fof}. If two particles are separated by less than a user defined linking length, the particles are assigned to the same group. This results in groups of connected particles, the so-called FOF groups. On top of those two methods, a large variety of algorithms has been built over the last two decades: a recent halo finder comparison paper \citep{knebe_state} listed 38 different halo finders. For more detailed information about the halo finders which are on the market today, we refer to the series of papers that has emerged from the halo finding comparison project \citep{haloes_mad, haloes_notts1, knebe_state, haloes_notts2}.

On even larger scales, the identification and characterization of cosmic voids is an important task. Similar to haloes, the voids assemble in a hierarchical structure of voids and sub-voids which can be found in observational and simulation data likewise. \cite{way2011structure} and \cite{way2014structure} give an overview on void finding techniques and the relation to the identification of overdensities.

Automatic detection of structure is also performed at galactic scales. For example, Astronomers performing radio observations of molecular clouds entered the field when they started to identify clumps in position-position-velocity (PPV) data cubes. \cite{Gaussclump} tried to fit the data by sums of triaxial Gaussian-shaped clumps and \cite{Clumpfind} identified structure by contouring the dataset at evenly spaced levels without assuming an a priori shape for the clumps. More recently, \cite{rosolowsky2008structural} showed how dendrograms can be used to exploit the hierarchy that naturally arises from contouring a PPV cube at multiple emission levels and used this technique to define substructures in molecular clouds. 

With such a large choice of astrophysical structure finding tools at hand, one might ask the question why there needs to be yet another one. The trigger for the development of a new analysis tool was our need for ``on-the-fly" structure finding in the astrophysical simulation code \citep{Teyssier2002}, in order to locate gas and/or dark matter clumps while the simulation is running. As pointed out in \cite{knebe_state} there is a general trend towards ``on-the-fly'' analysis for many reasons: most modern astrophysical simulations are performed on large computational infrastructure with distributed memory. The sizes of those simulations often exceed the total memory present in commonly used shared memory machines. The structure finding is therefore preferentially performed on the same machine that is running the simulation. Beyond that, the sizes of one single output of such simulations can quickly reach hundreds of GBs, up to several TBs. Storing many outputs for later post-processing is often not possible due to limited disk space, so that keeping only a catalogue of structure is the only viable solution.

Another reason for detecting structures while the simulation is advancing, is the possibility to couple the results of the halo decomposition to the simulation itself. In \cite{sink_paper}, for example, we have described a new algorithm for creation of sink particles, based on the properties of gas clumps detected ``on-the-fly''. This application requires an extremely high frequency at which structure finding must be performed. It must therefore make efficient use of the parallel infrastructure, and deliver good scaling properties for increasing numbers of MPI tasks, up to the number of CPUs the simulation is running on. Otherwise it will unacceptably slow down the simulation.

These requirements resulted in the development of \textsc{phew} (\textbf{P}arallel \textbf{H}i\textbf{E}rarchical \textbf{W}atershed), a new structure finding algorithm and its implementation into \textsc{ramses}\footnote{The \textsc{ramses} code including \textsc{phew} are publicly available and can be downloaded from \url{http://www.bitbucket.org/rteyssie/ramses}}. While \textsc{phew} is not based on any pre-existing algorithm, it combines various concepts that have been used in other astrophysical structure finding tools before. 

First, \textsc{phew} falls into the category of ``watershed-based'' algorithms. These algorithms assign particles or cells to density peaks by following the steepest gradient, resulting in the so-called ``watershed segmentation'' (see Section \ref{watershed}) of the negative density field. Other members of this category are \textsc{denmax} \citep{Bertschinger1991}, \textsc{hop} \citep{Eisenstein1998},  \textsc{skid} \citep{Stadel2001}, \textsc{adaptahop} \citep{Aubert2004}, \textsc{grasshopper} (Potter et al., in prep). Note that in contrast to the aforementioned codes which work on the particles directly, we use a mesh to define the density field\footnote{\textsc{denmax} can be considered an in-between case since it uses a uniform grid to compute the density gradient which is then used to directly assign particles to peaks.}. Void finding is typically performed using watershed-based algorithms too  \citep[e.g.,][]{platen2007cosmic,aragon2010spine,sutter2015vide}

Second, region merging in \textsc{phew} is based on the topological properties of saddle surfaces. This is the case as well for \textsc{hop}, \textsc{adaptahop} and \textsc{subfind} \citep{Springel2001}. As in the \textsc{ahf} halo finder \citep{knollmann2009ahf}, \textsc{phew} works on the density field deriving from particles that were previously projected onto the AMR mesh. In contrast to \textsc{ahf}, however, we do not use the AMR grid as a way of contouring the density field. A low density region which - for whatever reason - is refined to a high level does not compromise our results. Thus, in the landscape of existing halo finders, \textsc{phew} can be seen as filling the gap between  \textsc{p-hop} \citep{skory2010parallel} which does not find substructures but is a MPI-parallel version of \textsc{hop}, and \textsc{adaptahop}, a multi-threaded software that does find substructures, but has not yet been MPI-parallelized.

The aim of this paper is to present a new structure finding algorithm that: 1- can be applied to any density field defined on an adaptive grid, 2- is capable of detecting substructure, 3- is parallelized using the MPI library on distributed memory systems, and 4- is fast enough to be run at every time step of a simulation without significantly slowing down the calculation. What is not discussed in the present paper is an unbinding procedure for particles that are located inside the volume occupied by a certain halo but not gravitationally bound to it, as well as the subsequent computation of halo properties. These functionalities will be added to \textsc{phew} in the future. 
As briefly mentioned above, a previous version of \textsc{phew} has already been presented in \cite{sink_paper}. The algorithm described here differs from the previous one in the sense that it is now fully parallelized. This allows the algorithm to run now efficiently on thousands of CPUs and handle a complex topography with millions of density peaks and a rich hierarchy of substructures.  

The article is organised as follows: in Section \ref{working principle} we describe the serial version of the \textsc{phew} algorithm. In Section \ref{parallelization} we focus on the parallel implementation of the steps presented in Section \ref{working principle}. Section \ref{scaling} contains scaling experiments which demonstrate the efficiency of the parallelization. Finally, we summarise and discuss our results, presenting an outlook on possible future work  in Section \ref{conclusions}.

%%%%%%%%%%%%%%%%%%%%%%%%
%%%%%%%%%%%%%%%%%%%%%%%%
%%%%%%%%%%%%%%%%%%%%%%%%
% work principle
%%%%%%%%%%%%%%%%%%%%%%%%
%%%%%%%%%%%%%%%%%%%%%%%%
%%%%%%%%%%%%%%%%%%%%%%%%

\section{The \sc{phew} \rm algorithm}\label{working principle}

\begin{figure*}
	\includegraphics[width=\textwidth]{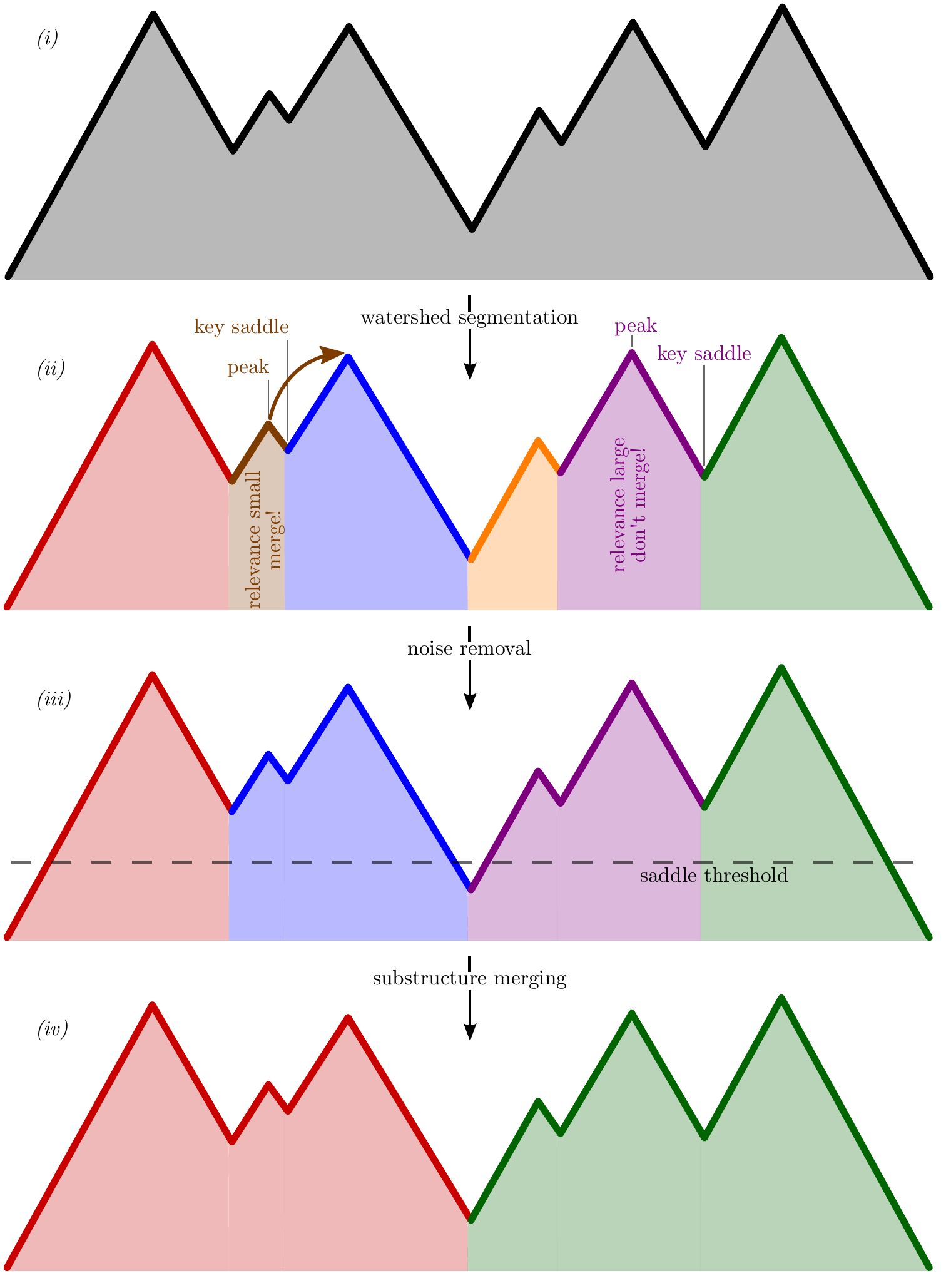}
	\caption{Visualization of the main steps of \textsc{phew} on a 1D density field (first panel). The segmentation into peak patches is shown in the second panel. Based on the relevance of a peak (peak-to-saddle ratio) we decide whether a peak represents ``noise'' or substructure. Irrelevant peaks are merged through their highest saddle points (third panel). The surviving objects are labeled as Level 0 clumps and denote the finest level of substructure. The substructure is merged based on a saddle threshold (third panel) into parent structure (fourth panel).}
	\label{soap}
\end{figure*}

In this section we describe the serial algorithm. As a starting point, we assume that we have a 3D density field on a AMR grid, particles have been projected onto the grid beforehand. 
The algorithm can be broken down in four main steps:
\begin{itemize}
\item Watershed segmentation
\item Saddle point search
\item Noise removal
\item Substructure merging
\end{itemize}
In the first step, we assign every cell above a user defined density threshold to a local density maximum by ascending along the steepest gradient. This results in a primary segmentation of the computational volume into ``peak patches'': regions associated to certain density peak. We establish the connectivity between the peaks by identifying the saddle points. We eliminate the peaks with a low density contrast to the background by merging them to a neighbour through their densest saddle point. The structure surviving the noise removal is considered the finest (sub)-structure. In a last step, we recursively merge the substructure to form larger and larger composite objects. 
 
\subsection{Watersheds in image processing}\label{watershed}

Before we start with a more detailed description of the algorithm, we take a quick look over the fence into the field of mathematical morphology and its application to image processing. There, watershed algorithms are a well known and extensively studied tool for image segmentation. The basic idea is that a grayscale image can be thought of as a topographic relief. A drop of water that falls somewhere onto this relief will follow the line of steepest descent until it reaches a local minimum. All points that connect to the same local minimum in that manner form a catchment basin. The watershed algorithm therefore segments the picture into catchment basins. The boundaries of the catchment basins are the actual watersheds. This technique is usually applied to the magnitude of the images gradient. In this way, the watershed lines trace regions of high gradients and segment the original image it into connected regions of small gradients. An excellent overview of the watershed techniques used in image processing is given by \cite{Roerdink2000}.

%\textcolor{red}{When comparing to watershed algorithms used for image segmentation, we have to consider a few aspects where our use case differs from the above one: some watershed algorithms are based on the image pixels being accessed in groups according to their gray level. While an 8 bit image contains only 256 gray levels, our density field is usually represented by an 8 byte float. Looping over all possible gray levels is clearly impossible in our case. Related to that, the limited number of grayscale levels introduces the problem of locally flat regions which do not contain a minimum. Since we use an almost continuous representation of densities we may safely ignore this issue. }

%\textcolor{red}{Another distinguishing feature of watershed algorithms is whether and how they construct watershed pixels as the boundaries between adjacent catchment basins. This is not important for us as we want to assign every boundary cell to one or the other catchment basin. In this philosophy, the cell surfaces are the actual watersheds which are constructed from the segmented density field after the actual watershed algorithm has finished.  }

An important difference to the watershed algorithms used for image segmentation lies in the computational cost for checking all neighbours of a cell/pixel. Working in 3D naturally increases the number of neighbours. Using an AMR grid further increases the number of possible neighbours since one has to consider possible neighbours at the same level as the original cell as well as one level above and below. Most importantly, the data structure in an AMR grid is very different from the one of a flat 2D array. The location of neighbouring cells in memory needs to be constructed before a neighbour can be checked for its density. Our main interest lies therefore in reducing the number of neighbours that have to be accessed. This aspect influences the choice of watershed algorithm for our purpose.

\subsection{Watershed segmentation}

In a first step, all cells above the density threshold are marked. We call those cells ``test cells''. For every test cell the densest neighbouring cell is identified and stored. If a cell has no denser neighbour, it is a local density peak. The peak obtains a peak ID which is stored as the ``peak patch label'' of the corresponding cell. The test cells are sorted by decreasing density. Once sorted, every cell copies the peak patch label from its densest neighbour. The previous sorting ensures that the densest neighbour has been accessed before and has therefore already obtained its peak patch label. Thus, every cell is assigned to a peak after this one pass. All cells marked with the same peak patch label form a peak patch (see Figure \ref{soap}, second panel). Note that our peak patches correspond to the catchment basins introduced in Section \ref{watershed}. Since we are working on peaks rather than minima, we introduce this new terminology to avoid the cumbersome notion of an ``inverted catchment basin''. Note that this procedure is very similar to the \emph{hill climbing} method described in \cite{Roerdink2000} which was introduced by \cite{meyer1994topographic}. 

\subsection{Saddle point search}\label{saddle point search}

Before we can merge peak patches, we have to establish the connectivity between them. All test cells are checked for neighbouring cells that belong to a different peak patch. If such a neighbouring cell is found, the average density of the starting cell and its neighbour is considered as the density at the common surface of the two bordering peak patches. The maximum density on the connecting surface is the density of the saddle between the two peaks and stored. At the end of this step, each peak has its list of neighbouring peaks together with the corresponding saddle point densities. We denote the maximum saddle point of a peak as the ``key saddle'' and the corresponding neighbour as ``key neighbour''. 

\subsection{Noise removal}\label{noise removal}

A known problem of the watershed method is over-segmentation. The presence of a huge number of local minima - for example due to random particle noise or transient gas density fluctuations - causes segmentation into as many catchment basins as there are local minima. Generally speaking, there are two possible strategies to deal with this problem: not creating the over-segmentation in the first place or merging over-segmented regions. Preventing over-segmentation the can be obtained using markers to preselect allowed minima \citep[e.g.,][]{moga1998marker}. This usually requires a human intervention, which in our case is not possible. Another way is to use the so-called hierarchical watershed algorithm\footnote{Note that more modern approaches to region merging in image segmentation use the original image for merging while the watershed is computed on the gradient image \citep[e.g.,][]{peng2011automatic}. Using the watershed on the gradient image results in regions of similar gray values, where the densities inside our peak patches are very inhomogeneous. Approaches to region merging are thus fundamentally different in image processing than they are in our case.} \citep{Beucher1994}. Hierarchical watershed algorithms merge artificial catchment basins to more important ones based on some criteria. What we will describe in the following turns our watershed algorithm into a hierarchical algorithm in the \cite{Beucher1994} sense.%\textcolor{red}{, where our merging criterion is inspired by the notion of a signal-to-noise ratio.}

After having previously identified the saddle points, we classify the peaks based on their contrast to the background. We define the contrast as the ratio of the peak density to the key saddle density and name it ``relevance''. This is sketched in the second panel of Figure \ref{soap}. Every peak is assigned a ``final peak'' label which is initialized to the peaks own peak ID and updated whenever a peak is merged to another one. The peaks are sorted by decreasing peak density. For each peak, the key saddle is determined from the list of saddle points and the relevance is computed. Peaks with a relevance below a relevance threshold are considered noise\footnote{The relevance threshold is a user parameter that can be adapted to the setup. 1.5 is our standard choice for identifying gas clumps in \textsc{ramses} simulations. For identifying dark matter haloes, the value can be picked according to the expected number of dark matter particles per cell and the resulting Poisson noise in the density.}. If the peak is relevant, it is not touched. For an irrelevant peak, we check whether its key saddle links it to a denser peak. If this is the case, it will inherit the final peak label from this key peak. As in the watershed segmentation, the previous sorting makes sure that the final peak labels can propagate through long chains of connected peaks in just one loop. If a peak is both isolated and irrelevant, it is discarded.

When two peaks merge, their lists of saddle points are merged as well. If both peaks used to have a connection to the same third peak, the maximum of the two saddles is kept. 

Now, we iterate the procedure: from the updated lists of saddle points, the key saddles are determined. Peaks are accessed in the order of decreasing peak density and irrelevant peaks are merged. After an iteration without any mergers, all irrelevant peaks have been merged or discarded and the noise removal is finished. Note that the described merging process follows exactly the same principle as the watershed segmentation. We have simply replaced cells by peaks, densest neighbour cells with key neighbours and the peak patch label by the final peak label. We call the structures which survive the noise removal {\bf Level 0 clumps}. They constitute the finest structure (see Figure \ref{soap}, third panel) in our hierarchy.

Using the relevance as a merging criterion results in a similar definition of a clump as it is obtained by algorithms that contour the dataset at evenly spaced levels in log-space \citep[e.g.][]{Clumpfind}. There, a peak-to-saddle ratio above a given value guarantees that a contour level will fall between peak density and key saddle density and thus the detection of the corresponding clump as an individual object. However, a contour level can coincidentally lie between the peak density and the key saddle density of an object with a very low peak-to-saddle ratio, resulting in the detection of an irrelevant density fluctuation as a clump. Merging based on the relevance removes this randomness from the analysis. 

For a density field that is obtained from an underlying particle distribution, the relevance criterion can be interpreted as a signal-to-noise criterion on the basis of an individual cell. We assume a roughly constant number of particles per cell as this number is often used as a refinement criterion for dark matter simulations in \textsc{ramses}. A large relevance thus translates into a small probability that the peak density is simply drawn from a Poisson distribution with the mean being equal to the saddle point density. A \emph{true} signal-to-noise criterion would consider the probability that the entire peak patch is consistent with being randomly drawn from the density at the saddle point. We would expect such a criterion to distinguish noise from physical structure more reliably. However, such a criterion is not compatible with our parallelization strategy of the merging procedure, as it includes quantities that are ``additive'' under a merger - such as the size or the total mass of a peak patch - into the merging criterion. As we will describe in Section \ref{merging order}, this would make the outcome of the merging process depend on the exact order at which the peaks are considered for merging.

\subsection{Saddle threshold merging}

If desired, the remaining peaks and their associated clumps can be merged further to form composite clumps. This happens by exactly repeating the previous merging process with a different merging criterion. We have implemented a density threshold for the key saddle as a criterion. If the key saddle density is above that threshold, a peak is merged to its key neighbour (see Figure \ref{soap}, fourth panel). Another possible criterion is the repeated use of the relevance threshold, this time with a higher value.

\subsection{A hierarchy of saddle points}\label{merging_order}

\begin{figure}
	\includegraphics[width=\columnwidth]{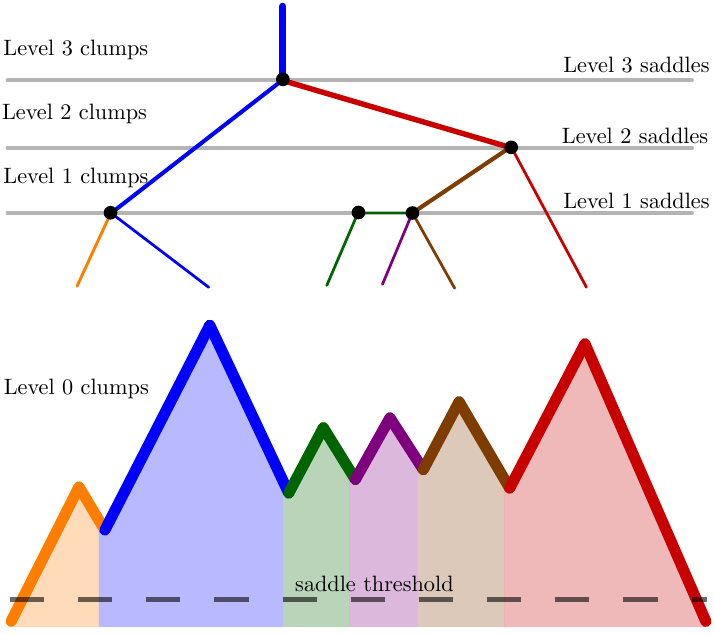}
	\caption{Hierarchy of saddle points as it is produced by our merging algorithm. Level $n$ saddle points are used for merging during the $n$-th round of mergers. Level $n$ clumps emerge from a merger through a Level $n$ saddle point and contain $n$ levels of substructure. }
	\label{hierarchy}
\end{figure}

We have seen in Section \ref{noise removal} that saddle points are removed in groups or levels by merging through them. All key saddles which link their peak to a denser one are removed at once. Through the merging, other saddle points become key saddles and the next level of saddle points is removed. By repeating this process, a natural hierarchy of saddle points and clumps is produced. In Figure \ref{hierarchy} we illustrate the construction of this hierarchy. We start with the Level 0 clumps after the noise removal (no substructure except for noise) and assume that the saddle threshold for merging is below any of the saddles depicted in Figure \ref{hierarchy}. The Level 1 saddle points are identified and used for merging. The resulting objects are Level 1 clumps as they have one level of substructure. In general, a Level $n$ clump is formed through a merger which removes a Level $n$ saddle point and contains $n$ levels of substructure. This produces a very natural hierarchy of saddle points and clumps based on the levels of substructure. Note that the level of a saddle point does not reflect its density. A more traditional way of grouping substructure based on the density of the saddle that connects two substructure objects as it is for example produced by \textsc{adaptahop} can easily be recovered from this hierarchy.

\subsection{Merging order}\label{merging order}
We will see in Section \ref{parallelization} that we have to drop the idea of sorting the peaks globally when we parallelize \textsc{phew}. This will alter the order in which peaks are merged in an unpredictable way. It is therefore crucial that the \textsc{phew} allows the order of mergers to change without causing different results. This not true in general. Yet, as we will show in this section, it is the case when we respect the three merging rules:

\begin{enumerate}
\item A peak is only merged to a denser one (upward).
\item A peak is only merged through its key saddle.
\item The density of the key saddle or the relevance are used as merging criterion.
\end{enumerate}

The result of the merging procedure is uniquely determined by the set of saddle points that is used for merging. This is a subset of all saddle points. In order to affect the outcome of the merging process, changing the order of mergers therefore has to change the set of used saddle points. Let us consider a peak $n$ connected to its key neighbour $m$ through the key saddle $s_{nm}$ at the very beginning of the merging process. The peak density of $m$ is higher than that of $n$, $m>n$. There are three possible types of mergers related to $n$ or $m$ that can happen before $n$ is considered for merging. We will show that none of them can change the fate of $n$.

\begin{enumerate}
\item A third peak might be merged into $m$. Due to upward merging, this cannot change the peak density of $m$ and therefore  decision if $n$ will be merged into $m$ is not influenced. 

\item Peak $m$ might merge into another peak $m'$. The saddle $s_{nm}$ will still exist, now linking $n$ to $m'$. Due to upward merging we have $m' > m > n$ which means that $n$ is still the lower of the two peaks connected by $s_{nm'}$. The decision whether $n$ is merged through $s_{nm}$ is unaltered.

\item A third peak $i$ might be merged into $n$. The peak density of $n$ cannot change due to that since it would mean that peak $i$ had a higher density than $n$ which contradicts the upward merging. The key saddle cannot change because this would mean that peak $i$ had a saddle point $s_{ij}$ higher than $s_{nm}$. This would imply that the saddle point $s_{ni}$ through which $i$ was merged into $n$ was even higher, $s_{ni} > s_{ij}$ otherwise $s_{ni}$ had not been the key saddle of peak $i$. Yet, $s_{ni} > s_{ij}>s_{nm}$ contradicts that $s_{nm}$ is the key saddle of peak $n$. The peak density of $n$ and its key saddle are thus unchanged, therefore the relevance of $n$ is not changed either.
\end{enumerate}

This shows that we can arbitrarily delay the moment when we consider a peak for merging as long as we respect the three merging rules. The mergers happening in the mean time cannot change the properties deciding if and through which saddle this peak will be merged. A possible way to prevent violation of merging rule (ii) is to consider all peaks for merging until no further mergers are possible before any new key saddle of the merged peaks is computed. This results in using the saddle points for merging on a ``level-by-level'' basis. This is a key to the parallelization of \textsc{phew} since it will allow performing a big number of operations (mergers), in between each round of communication (finding new key saddles).
Note that this line of argumentation breaks when we violate merging rule (iii) and use for example the clump mass as merging criterion. The mass is a property that changes with every merger. Therefore, altering the merging order does change the mass of a clump at the moment it is considered for merging and can thus change the decision whether the clumps should be merged or not.

\section{Parallel implementation}\label{parallelization}

We now turn to the implementation of the previously described steps in a parallel, distributed-memory framework. Where a detailed description of an algorithmic block in words would prevent readability of the paper, we refer the interested reader to a corresponding block written in pseudocode located in Appendix~B. We assume that the computational domain has
been previously decomposed into non-overlapping spatial domains, each domain containing a partition of the AMR mesh on which the density field is defined. In every MPI task, the local partition of the mesh is referred to as the ``active cells''. They are wrapped by a thin layer of cells that belong to other tasks. These ghost cells are referred to as belonging to the ``virtual boundaries''. These virtual boundaries are updated through MPI communication before \textsc{phew} is called to make sure that the densities in the virtual boundary cells are equal to the densities in the corresponding active cells hosted by other MPI tasks. 

\subsection{Parallel watershed}\label{parallel watershed}

The watershed segmentation is \emph{non-local} by nature. This can easily be understood by imagining a mountain ridge. Two drops of water falling onto both sides of the ridge will initially move away into different directions. They might flow into different rivers which flow into different lakes, or they might as well end up in two rivers which join before reaching a lake. The two situations \emph{cannot} be distinguished based on local properties. Parallelization of the watershed algorithm is therefore a non-trivial task. In the literature, one finds various approaches to parallelization for the different watershed algorithms \citep[see e.g.][]{Roerdink2000}. Our technique is very close to the technique described in  \cite{moga1997parallel} 
and called ``hill climbing by locally ordered queues''.

\begin{figure}
	\includegraphics[width=\columnwidth]{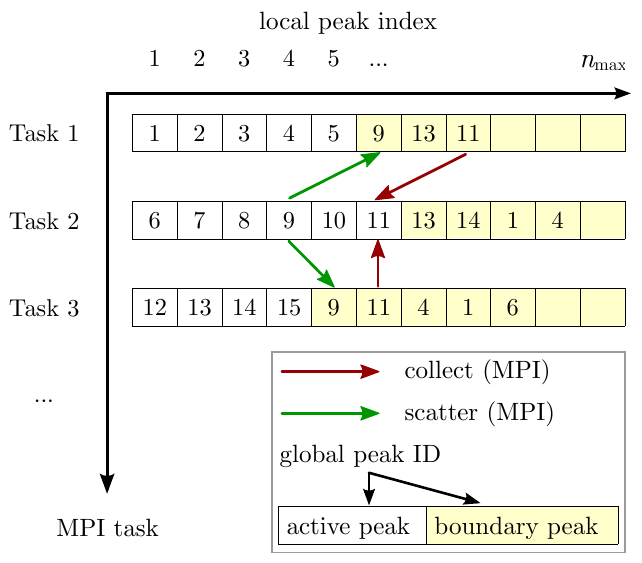}
	\caption{Example of peak layout in memory for 3 MPI tasks. The figure shows the global peak ID as a `function' of the MPI task and the local peak ID. The local peak index for a given global peak ID is stored in a hash table.}
	\label{pmemory}
\end{figure}

\begin{figure}
	\includegraphics[width=\columnwidth]{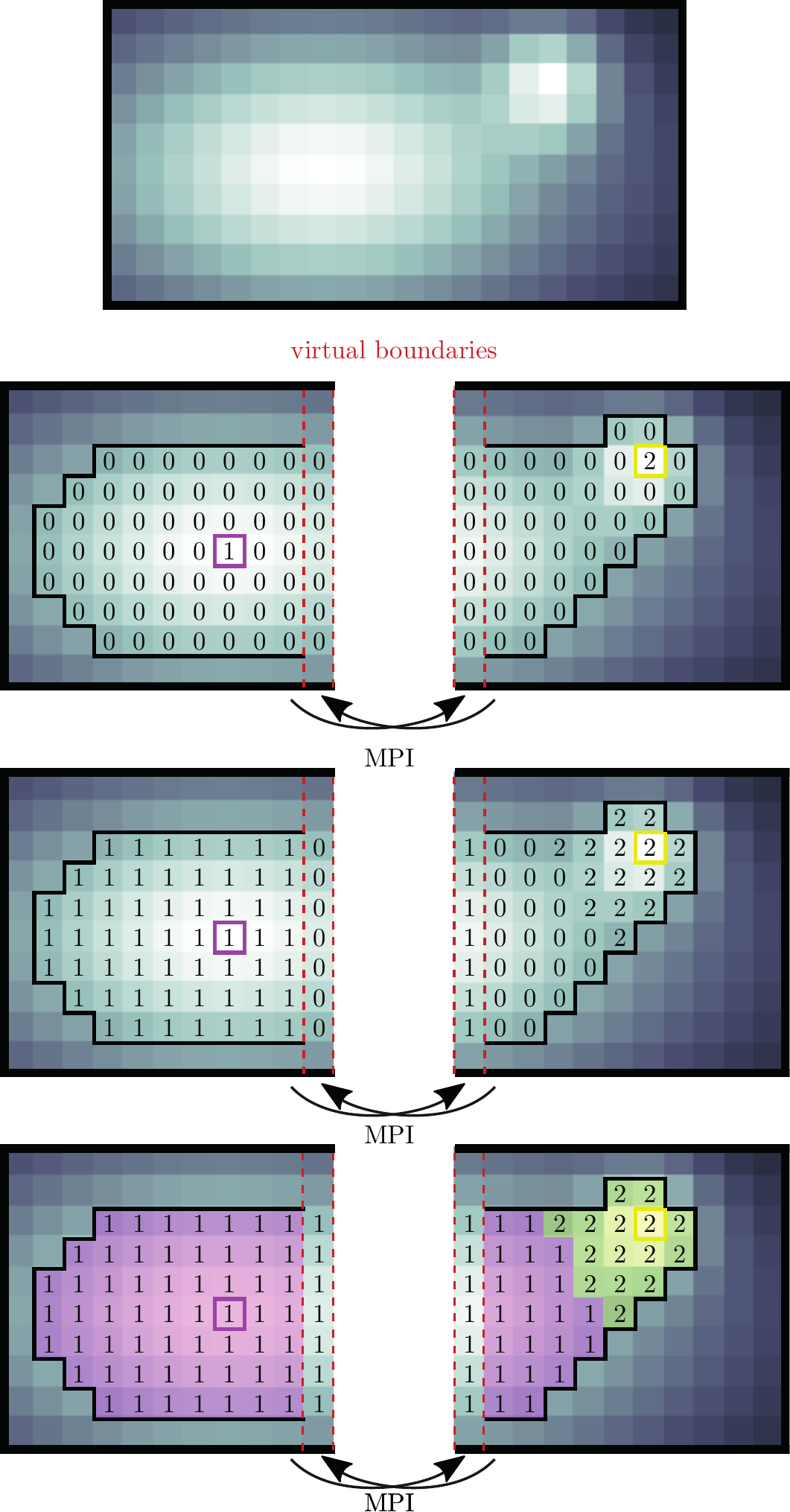}
	\caption{Parallelization of the watershed segmentation shown on a 2D field. The top panel depicts the computational box with the density field. In the second panel, the two MPI domains and the virtual boundaries are shown, the peaks have obtained their IDs and the cells are labeled. In a loop over all test cells, the peak patch labels can propagate inside the MPI domains (third panel). After the loop, the virtual boundaries are updated and the procedure is repeated (fourth panel). }
	\label{ppatch}
\end{figure}

Each task performs a loop over all its active cells, in order to identify first the test cells (cells above the density threshold). For faster access, the indices of all test cells are stored in an array. A loop over all test cells is performed where the densities of all neighbouring cells are checked. The index of the densest neighbouring cell is stored for each test cell, since it will be used several times during the algorithm. Note that the densest neighbour of a cell can lie inside the virtual boundary, while test cells are always inside the active domain. 

During the first loop, all peaks (local extrema) are counted. After the loop, the number of peaks in each MPI domain are communicated between all MPI tasks, which allows each MPI task to compute a global index (ID) for its peaks (see Figure~\ref{pmemory}). In another loop over test cells, cells which represent a peak are labeled with their global peak ID, all other test cells are initialised with a peak patch label equal to zero. The peak patch labels are updated inside the virtual boundaries using MPI communication (Figure \ref{ppatch}, second panel). As explained in Section~\ref{working principle}, every MPI task computes a permutation which sorts test cells in decreasing density order, using the quick sort algorithm \citep{NR}. Using this permutation, a sorted loop, where every cell inherits the peak patch label from its densest neighbour is performed (Figure \ref{ppatch}, third panel). During this loop, the number of cells that have changed their peak patch label is counted. After the loop, the peak patch labels in the virtual boundaries are updated again through MPI communications. This procedure is iterated (Figure \ref{ppatch}, fourth panel) until no cell inside the entire computational box has changed its peak patch label during a full loop. This completes the parallel watershed segmentation.

\subsection{Virtual peak boundary}

As we have already described in Section \ref{working principle}, our peak patch merging step is analogous to the  segmentation step. 
The patches now take the role of the cells, the peak patch label is replaced by the final peak label and the densest neighbouring cell is replaced by the key neighbouring patch. 
As explained before, the parallelization of the peak patch segmentation is exploiting the virtual boundaries surrounding each MPI domain. If we want to use the same strategy to parallelize the merging process, we  need the analog of the virtual mesh boundary: a virtual peak boundary. In contrast to our usual virtual mesh boundary, the virtual peak boundary does not represent a fixed region in space. As the merging process advances, new connections appear and new peaks have to be introduced in the virtual peak boundary. Our virtual peak boundary is therefore more dynamic than our virtual mesh boundary. 

Figure \ref{pmemory} shows a possible layout of peaks in memory. Note the distinction between a peaks global ID and its local index. The latter of the two is the position of the peak in local memory. The peaks that are located inside a tasks MPI domain are called active peaks. They take the first $N_{\rm active}$ places in memory. The active peaks are followed by the ghost peaks that belong to the virtual peak boundary. Since it is unknown at the beginning of the merging process how much space for ghost peaks will be necessary, we set 
\begin {equation}\label{theonlyone}
N_{\rm max}=\max\{4\max\limits_{\rm tasks} \{N_{\rm active}\},1000\}, 
\end{equation}
as a default value that can be modified by the user. The preset $N_{\rm max}$ is mostly a large overestimation of the effectively used space in memory for peaks (see, fourth row in Table \ref{parastats}), designed to be sufficient for all setups we have tested. However, the memory consumption for peak properties is still negligible compared to the necessary space for the AMR grid.\footnote{For situations where the memory consumption due to given estimate for $N_{\rm max}$ becomes prohibitive, one could start with a lower number and for example double the size of the allocation on-the-fly whenever all available space for ghost peaks is occupied. However, we have not yet encountered a situation where it was necessary to use this strategy.} All peak properties such as the peak density are allocated up to $N_{\rm max}$.

Since every task is aware of its starting number of global peak IDs, switching from global peak ID to local peak index and vice versa is trivial for active peaks. To recover a boundary peaks global ID from its local index, we simply store the global ID in memory at the position of its local index. For the opposite direction we use a hash table that contains the local peak index for a given global peak ID (hash key)\footnote{We use a simple hash function based on the remainder of a division of the peak ID by a prime number chosen according to the maximum size of the virtual peak boundary. Collisions are dealt with by chaining in the form of a linked list \citep{Knuth}. We found this to be sufficient for our purpose (see Table \ref{parastats}).}. Whenever we introduce a new boundary peak into the virtual peak boundary, it obtains the local peak index corresponding to the first free space in memory. The global peak ID is stored and a hash key is computed.
Which peaks need to be present in the virtual peak boundary depends on the connectivity of peaks. The initial state of the virtual boundary will thus be constructed while searching for saddle points that connect the peaks.

\subsection{The peak communicator}

By introducing a peak into the virtual peak boundary, it only obtains a local peak index. No properties except the global peak ID of a newly introduced boundary peak are present at this stage. We now describe how information is transferred from the MPI task which hosts a peak (the ``owner'' of that peak) into the virtual peak boundaries of other tasks and vice versa. There are two types of communication: inward communication (``collect'', red arrows in Figure \ref{pmemory}) from all processes which have a certain peak inside their peak boundary to the owner of the peak, and outward communication (``scatter'', green arrows in Figure \ref{pmemory}) to update the peak properties in the virtual boundaries. When performing a collect communication, one has to specify whether one is computing a sum, minimum or maximum of the incoming values belonging to the same peak. When a scatter communication is performed, the peak properties of boundary peaks are overwritten with their equivalent from the peaks owner. A typical communication pattern for a peak property is therefore a collect communication followed by a scatter communication. 

Before this communication can  be performed, we need to build a communication structure which we refer to as the ``peak communicator''. We allocate a matrix $C$ of size $N_\text{task} \times N_\text{task}$. The entry $c_{ij}$ is the number of peaks inside the virtual peak boundary of task $i$ that are owned by task $j$. Each task builds its line of $C$ in a loop over the boundary peaks by looking at their global peak IDs. Through MPI communication, the lines of $C$ are shared between all task to complete $C$\footnote{The introduction of a $N_\text{task}^2$ sized matrix can become problematic when the number of MPI tasks is increased beyond the numbers we have tested for this publication, especially for supercomputers with relatively little available memory per core (Blue Gene architecture). In order to apply \textsc{phew} to even larger problems, one can drop the construction of the global matrix $C$ by exploiting the fact that the $n$-th MPI process only needs to be aware of the $n$-th row and the $n$-th column of $C$, but not of the entire matrix. Considering the fact that the rows/columns of $C$ are sparse, one can thus replace the $N_\text{task}^2$ sized matrix by a fully scalable representation of the information contained in $C$.}. The entries in the matrix $C$ determine the amount of data that is sent to/received from another MPI task. This information is used to allocate send and receive buffers and to direct each entry in a tasks send buffer to the correct MPI task in a round of all-to-all communication. In order to complete the setup of the peak communicator, we use the established structure to perform a collect communication of the global peak ID. This information allows the identification of a position in the receive buffer (or in the send buffer in the case of a scatter communication) with an active peak. This completes the buildup of the communication structure. The peak communicator needs to be rebuilt whenever new peaks have potentially been added to the virtual peak boundary of any MPI task.

\subsection{The saddle point matrix}

To keep track of the saddle points, we establish a symmetric saddle matrix $M$, where the entry $m_{ij}$ is the density of the saddle point connecting the peaks $i$ and $j$. As most of the peaks patches are not touching each other, we use a sparse matrix representation of $M$. Note that the indices $i,j$ are the local peak indices, which makes $M$ a sparse matrix of virtual size $N_\text{max} \times N_\text{max}$. Since we are interested in the maximum entry of each line and the column where it is located when in comes to merging, we keep track of those two values when adding new entries into $M$. The maximum and its column need to be recomputed by checking each non-zero element of a line only after values have been removed from the given line in $M$ which reduces the number necessary accesses to the sparse matrix.

The construction of the sparse matrices is performed locally the way described in Section \ref{saddle point search}. Whenever a connection is found to a peak that is not yet present in the virtual peak boundary, the given peak is introduced by assigning it a local index. See Algorithm \ref{algo: saddle point search} for the pseudocode describing the saddle point search on each task. 

\subsection{Communication of saddle points}

We could now use a collect communication on the saddle points for every peak in the entire computational box. As a result of that, every task would have access to all saddle points of all his active peaks. The global key saddle and key neighbour could then be determined by every MPI task for his active peaks. However, this approach would introduce a lot of communication and unnecessarily fill the sparse saddle matrices. The only necessary information to perform one iteration in the merging process is the (global) key saddle density of a peak and the corresponding key neighbour. This global maximum saddle can be found by comparing the local maxima of each MPI task. We thus minimise communication by performing a collect communication only on the local maximum of each row in the saddle point matrix. Together with the local maximum saddle density, we collect the global peak ID that denotes the local key neighbour. The owner of a peak can now compute the global key saddle for a given peak by comparing all the local maxima. The global peak ID that was received from the MPI task which hosts the global key saddle is the key neighbour of the peak. If not already present, the key neighbour is introduced into the virtual peak boundary of the owner task and the key saddle density is written into the sparse saddle matrix of the owner. Every MPI task can now perform a complete iteration in the merging process without any further communication of saddle point densities.

\subsection{Merging in parallel}

We are now set for the actual merging of the peaks. We introduce two new peak properties: a logical variable called {\ttfamily alive} which is initialised to ``true'' and set to ``false'' when a peak is merged into another one, and the final peak label which is initialised to the global peak ID for all active peaks. These two new properties and the peak density are updated in the virtual peak boundaries using a scatter communication. A permutation which sorts the active peaks by decreasing density is computed. Now we propagate the final peak label through the key saddles in a level-by-level fashion. On each level, we iterate until no final peak label is moved, while the virtual boundaries are updated after every iteration. This is perfectly analogous to the parallel watershed segmentation. After every level of saddle points we update the {\ttfamily alive} variable, the saddle point matrices and the virtual boundaries. The merger routine is described in Algorithm~\ref{mergeralgo} in pseudocode. The substructure merging is performed in exactly the same way, we just replace the relevance threshold by the saddle density threshold.

\section{Scaling test}\label{scaling}

\begin{figure*}
\includegraphics[width=0.99\textwidth]{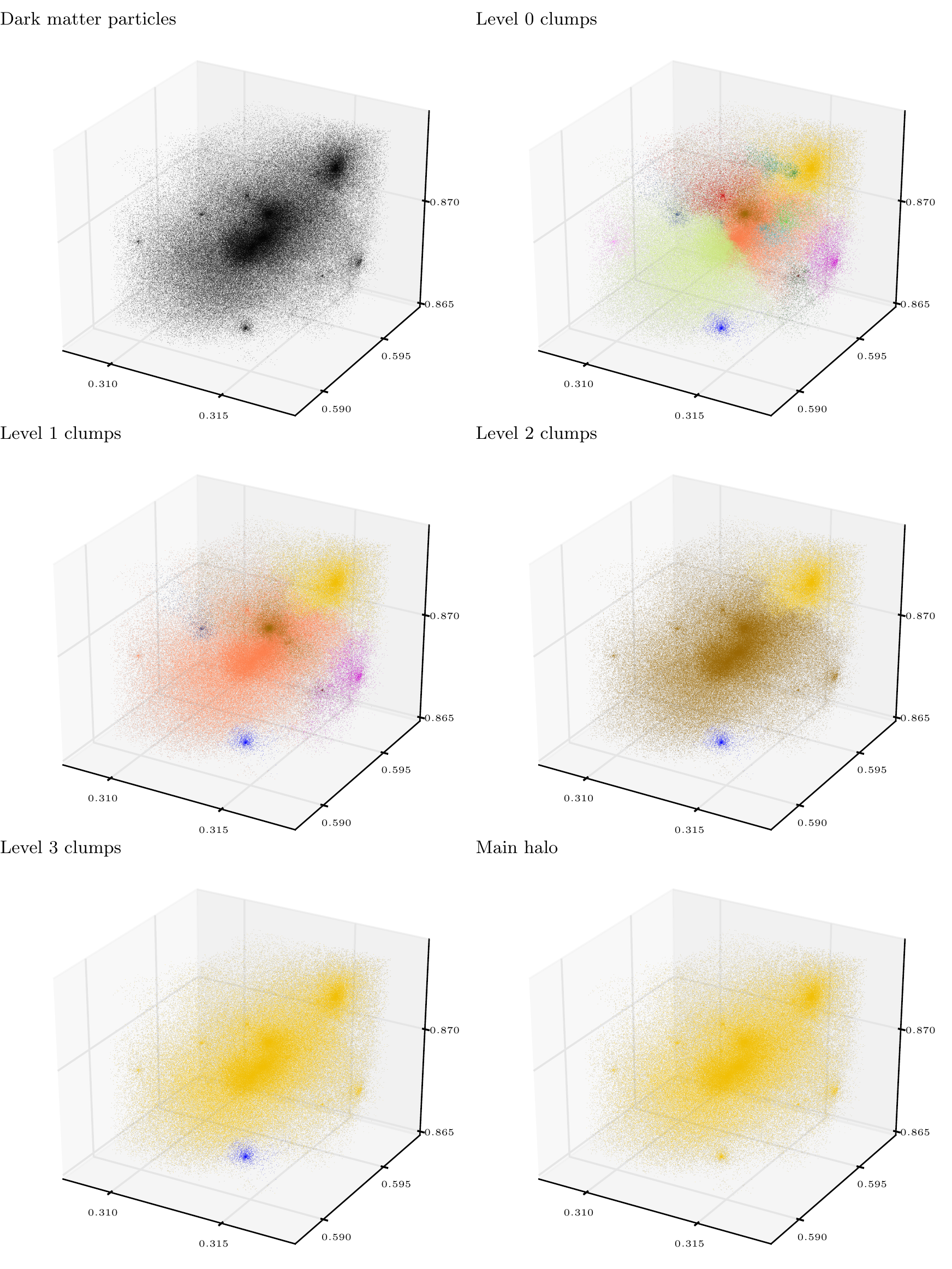}
\caption{}
\label{scatter}
\end{figure*}

We use a previously run cosmological dark matter simulation with $512^3$ particles for a scaling experiment. We restart the simulation from the output corresponding to redshift $z=0$ using various numbers of MPI tasks. Before \textsc{phew} can run, we project the particle density onto the AMR grid using the CIC \citep[Cloud-In-Cell,][]{Hockney1981} algorithm. 
Once we have constructed the grid-based density field, we run \textsc{phew} with a density threshold of 80 times the cosmological critical density (noted $\rho_{\rm crit}$) and a relevance threshold of 3. After merging the peak patches into Level 0 clumps (sub-haloes), we merge to form haloes by applying a saddle threshold of $200 \rho_{\rm crit}$. The first column in Table \ref{1024stats} summarizes parameters and runtime statistics obtained for 1024 tasks. We see a rich hierarchy of saddle points spread over many levels. The numbers of iterations necessary show that there is structure extending over several domain boundaries at every stage of the process (peak patches, clumps, haloes). Note that we \textsc{phew} finds \emph{exactly} the same structures, independent of the number of MPI tasks that have been used. This empirically confirms what we have described in Section \ref{merging order}. It is also worth mentioning that the iteration pattern looks surprisingly similar for the other $512^3$ runs in our scaling experiment. The total number of necessary iterations increases from 35 to 45 when going from 32 to 2048 tasks while it would be only 3 when for the serial algorithm.
An example of the hierarchical structure that is found by \textsc{phew} is shown in Figure~\ref{scatter} which depicts a halo with four levels of substructure taken from our scaling experiment. 
\begin{table}
	\caption{Parameters and some runtime statistics for the 1024 task runs of the experiment.}	
	\label{1024stats}	
	\renewcommand{\arraystretch}{1.2}
	\begin{tabularx}{1.0\columnwidth}{ L{0.45\columnwidth} L{0.225\columnwidth}L{0.225\columnwidth}}
	\toprule
	$N_{\rm parts}$				&$512^3$ 			&$1024^3$\\	
	$N_{\rm tasks}$				&1024				&1024	\\	
	Density threshold				&$80\,\rho_{\rm crit}$ 	&$80\,\rho_{\rm crit}$\\
	Relevance threshold			&3					&3\\
	Saddle threshold				&$200\,\rho_{\rm crit}$	&$200\,\rho_{\rm crit}$\\
	
	\midrule										
	 Number of test cells 		&104\,360\,968& 835\,609\,288\\		
	 Number of density peaks  	&6\,714\,764& 53\,994\,995\\
	 Number of relevant clumps  	&1\,311\,208& 10\,612\,079\\
	 Number of haloes\footnote{Note that we do only count the objects that contain more than N dark matter particles.}  			& 521\,185& 4\,234\,746\\		
	 Runtime    					&8.0\,s   & 38.9\,s \\
	\midrule									
	Number of iterations for... \\
	...watershed segmentation & 7 						&9\\	
	...noise removal  \\
	Level 1		&7&		7			\\
	Level 2		&5&		6			\\
	Level 3		&4&		4			\\
	Level 4		&2&		3			\\	
	Level 5		&1&		2			\\	
	Level 6		&1&		1			\\
	Level 7		&1&		1			\\
	Level 8		&&		1			\\
	...substructure merging \\
	Level 1		&4		&3			\\
	Level 2		&3		&4			\\
	Level 3		&3		&3			\\
	Level 4		&2		&2			\\	
	Level 5		&1		&2			\\	
	Level 6		&1		&1			\\
	Level 7		&		&1			\\
	\bottomrule			
	\end{tabularx}\\
	\smallskip
\end{table}

\begin{figure}
	\includegraphics[width=1\columnwidth]{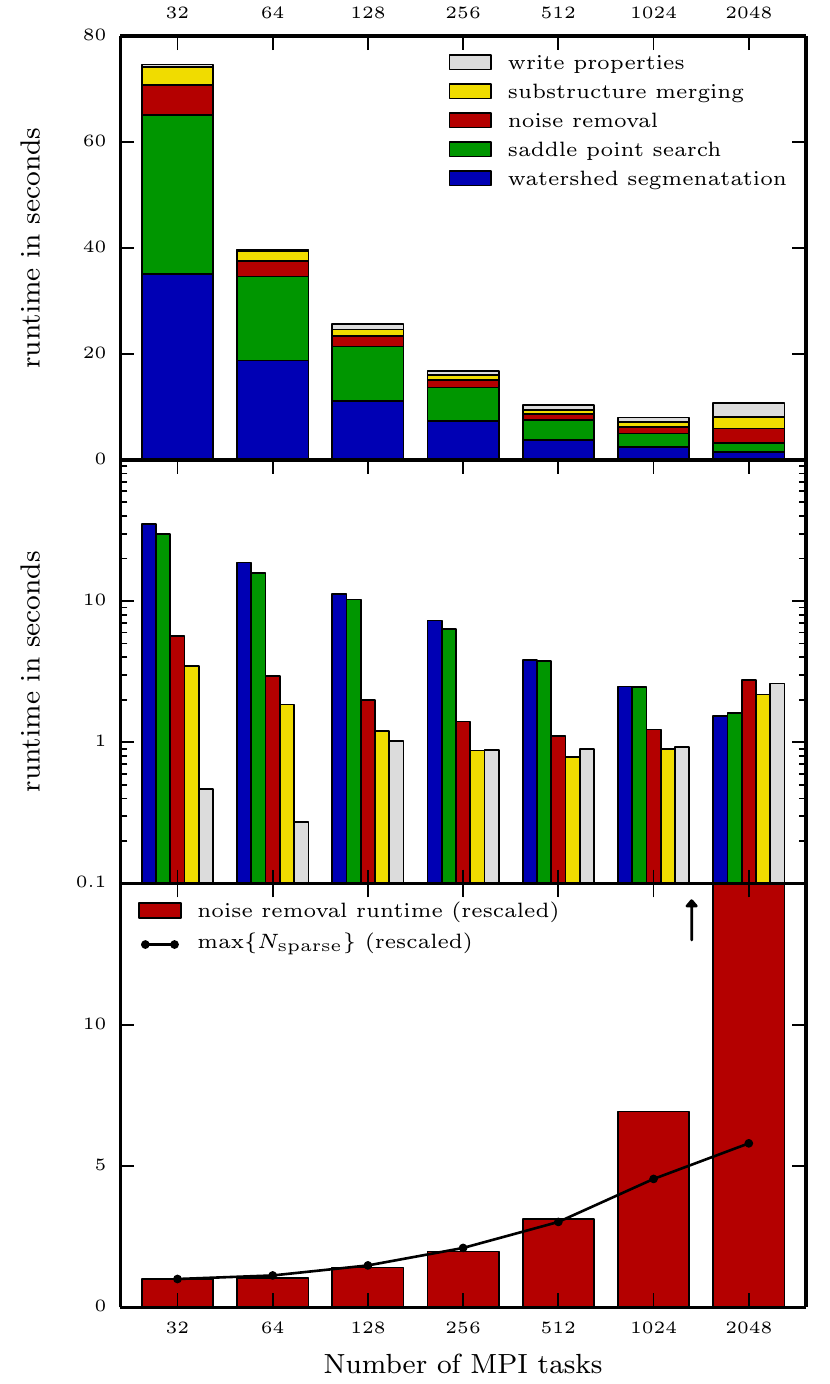}
	\caption{Scaling properties of the different parts in \textsc{phew} obtained by restarting a cosmological dark matter simulation with $512^3$ particles at redshift $z=0$. The top two panels show the runtimes of the different algorithmic blocks in \textsc{phew}. The peak patch segmentation and the saddle point search exhibit excellent scaling in the entire range of MPI tasks that we have tested. The merging in our test scales well up to $\sim 256$ MPI tasks. The bottom panel shows the maximum number of sparse matrix elements over all MPI tasks compared to $1/N_{\rm tasks}$ and rescaled to one at 32 MPI tasks. The increase seen in this number for of tasks is due to the growing load imbalance in terms of peaks per task and the increase in the surface to volume ratio of the domain segmentation. It explains the increase of the scaled runtime of the noise removal very well up to 512 tasks. The overall scaling of the algorithm is satisfactory up to 1024 MPI tasks which is four times the number of CPUs the original simulation was run on.}
	\label{scalingfig}
\end{figure}

\begin{table*}
	\caption{Runtime diagnostics for the parallelization of \textsc{phew} when various numbers of MPI tasks are used. $N_{\rm active}$ and $N_{\rm ghost}$ are the number of active peaks and ghost peaks respectively and $N_{\rm tot}=N_{\rm active}+N_{\rm ghost}$ denotes the total number of peaks per MPI task. $N_{\rm sparse}$ is the number of entries in the sparse saddle matrix and $N_{\rm collisions}$ gives the number of hash table collisions. Sums, maxima and averages are taken over the all MPI tasks. }	
	\label{parastats}	
	\renewcommand{\arraystretch}{2.0}
	\begin{tabularx}{1.0\textwidth}{ L{0.25\textwidth} L{0.08\textwidth} L{0.08\textwidth} L{0.08\textwidth} L{0.08\textwidth} L{0.08\textwidth} L{0.08\textwidth} L{0.08\textwidth}}
	\toprule
	
	$N_{\rm tasks}$		&	32&				64&				128&			256&			512&			1024&			2048\\	

	\midrule

	Load imbalance $\bigg(\frac{\max \{ N_{\rm tot} \}}{  \mathrm{avg} \{N_{\rm tot}\}} \bigg)$&
	
				 			1.4&				1.5&				1.8&				2.4&				2.8&				3.3&				3.9\\

	Surface effect $\bigg(\frac{\sum N_{\rm ghost}}{\sum N_{\rm active}}\bigg)$&	

							0.0087&			0.012&			0.016&			0.021&			0.030&			0.040&			0.055\\

	Connectivity $\bigg(\frac{\sum N_{\rm sparse}}{\sum N_{\rm tot}} \bigg)$&		 
							9.4&				9.4&				9.4&				9.3&				9.3&				9.3&				9.2\\

	$\max \bigg\{ \frac {N_{\rm ghost}}{N_{\rm active}}\bigg\}$&	 
	
							0.012&			0.017&			0.044&			0.064&			0.10&			0.15&			0.24\\

	$\max\{N_{\rm tot}\}$		 &	$3.0\times10^5$&	$1.6\times10^5$&	$9.6\times10^4$&	$6.4\times10^4$&	$3.8\times10^4$&	$2.2\times10^4$&	$1.3\times10^4$\\

	$\max\{N_{\rm sparse}\}$	 &	$3.3\times10^6$&	$1.8\times10^6$&	$1.2\times10^6$&	$8.7\times10^5$&	$6.3\times10^5$&	$4.7\times10^5$&	$3.0\times10^5$\\

	$\max\{N_{\rm collisions}\}$	
						 &	4&				3&				2&				3&				16&				17&				13\\

	\bottomrule			
	\end{tabularx}\\
	\smallskip
\end{table*}

In our numerical experiment, \textsc{phew} was run five times in a row, for five main simulation time steps following the restart. 
We measure the total runtime of each call to \textsc{phew} as well as the time spent on the different algorithmic steps. We find the variance of the runtimes to be negligible and conclude that the timings are stable. Note that the preliminary construction of the density field is performed inside the watershed segmentation block. However, the CIC algorithm is quick compared to the watershed segmentation. We also measure the amount of time necessary for each MPI task to write the properties of the structure inside its domain to disk.

The runtimes for the various numbers of MPI tasks are plotted in the top two panels of Figure \ref{scalingfig}. The top panel shows satisfactory scaling of the overall algorithm up to 1024 MPI tasks which is four times the numbers of tasks that were used to perform the original simulation. In this regime, the total runtime of \textsc{phew} is dominated by the watershed segmentation and the saddle point search. The most costly operations inside those two blocks are the construction and access of neighbouring cells. The total workload of those blocks thus scales linearly with the number of test cells per MPI task. 

The second panel shows that the runtime of those two blocks does actually scale over the entire range of numbers of tasks that we have tested. The second panel in Figure \ref{scalingfig} shows that the merging procedures scale well up to 256 tasks. The scaling of the merging process in this region is mainly controlled by two effects: with a growing number of tasks, the load imbalance of the peaks between the different MPI tasks increases. This is unavoidable as the domain decomposition is optimised for all AMR cells, not for the test cells only, and even less for the peak patches. The second reason is the growing ratio of surface to volume as the computational box is divided in smaller parts. This results in more ghost peaks per active peak which causes a higher workload per active peak. Those two effects are quantified in the first two rows of Table \ref{parastats}. 

The solid line in the bottom panel of Figure \ref{scalingfig} is a result of both effects mentioned above. It depicts $\max\{N_{\rm sparse}\}$, the maximum number of used sparse matrix elements over all MPI tasks. In perfect scaling conditions, this number would decrease as $1/N_{\rm tasks}$. We thus multiply $\max\{N_{\rm sparse}\}$ by $N_{\rm tasks}$ and rescale to one at 32 tasks. We compare this to the runtime of the noise removal (also scaled). We observe that this ``worst case'' number of entries in the sparse saddle point matrix does explain the scaling of the merging process up to 512 tasks. Beyond that, we believe that MPI communications become the performance bottleneck.

In Table \ref{parastats} we also show the maximum ratio of ghost peaks to active peaks. For 2048 tasks we have a value of 24\%. This shows that the number $N_{\rm max}$ defined in Equation \ref{theonlyone} is an overestimation of the effectively used memory for ghost peaks for this setup. In the same table, we also list the number of hast table collisions. There are very few collisions as the hash table is far from filling up and we conclude that the relatively simple hash function that we use is good enough for our purpose. Another fact worth mentioning is the relatively constant ratio of non-zero entries in saddle point matrix to the number of peaks seen in the third line of Table \ref{parastats}. Divided by two (due to the symmetry of the saddle point matrix), this number gives a good idea of the effective number of neighbours per peak.

\begin{figure}
	\includegraphics[width=1\columnwidth]{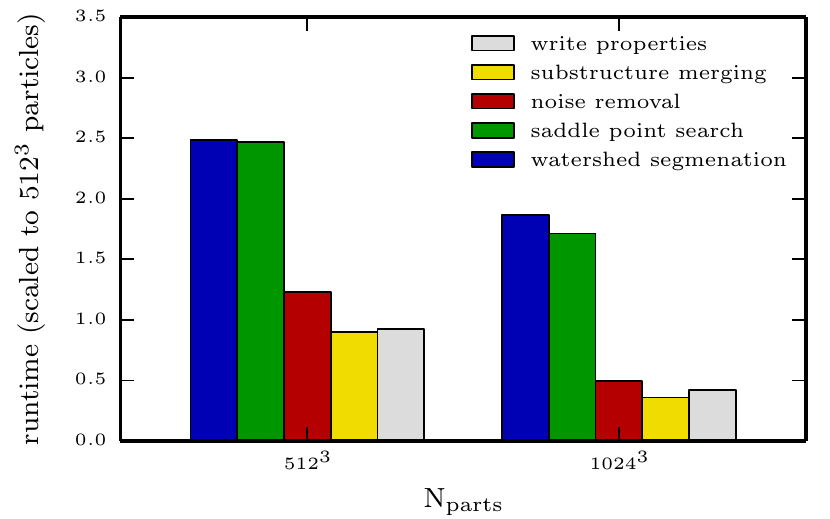}
	\caption{Weak scaling comparison of \textsc{phew} using 1024 tasks to find structure in a $512^3$ and a $1024^3$ particle cosmological box. The \textsc{phew} runtimes for the $1024^3$ box are divided by a factor of 8 for comparison with the runtimes for the $512^3$ box. Increasing the size of the dataset improves the scaling of \textsc{phew} for large numbers of MPI tasks.}
	\label{scalingfig_weak}
\end{figure}

As a second test we perform a ``weak scaling'' comparison of our 1024 task run with another 1024 task run but this time on a larger,  $1024^3$ particle box. The second column of Table \ref{1024stats} lists the statistics of that run. The numbers of test cells, peaks, clumps and haloes all increase by the expected factor of $\approx 8$. We thus divide the runtimes of \textsc{phew} for this setup by 8 and compare to the runtime of the 1024 task run on the $512^3$ box. This comparison is plotted in Figure \ref{scalingfig_weak}. The figure shows that the runtime per data decreases for all parts of \textsc{phew} by increasing the size of the data. Especially the efficiency of merging routines benefit a lot from the increased size of the dataset. We thus conclude that we can enlarge the range of $N_{\rm task}$ where \textsc{phew} scales well, by increasing the size of the simulation.

\addtocounter{figure}{-1}
\begin{figure}
\caption{Visualization of \textsc{phew} applied to a dark matter halo. We show a small sub-volume of the $512^3$ particle box used in our scaling experiment. The coordinates indicate the fraction of the box size. The sub-volume contains $\approx 2 \times 10^6$ particles. The objects that emerge after the noise removal (Level 0 clumps) are indicated in the second panel, where all particles belonging to the same object share a color. Every subsequent panel shows the status after a further round of merging as it is described in Section \ref{merging_order}. }
\end{figure}

\section{Conclusions}\label{conclusions}

We have presented \textsc{phew}, a new structure finding algorithm and its MPI parallel implementation into the AMR code \textsc{ramses}. \textsc{phew} finds density peaks and their associated regions in a 3D density field by performing a watershed segmentation. The merging is based on the saddle point topology. We have described a two-step approach to merging. 
In a first step, we merge irrelevant density fluctuations which we consider as noise. In a second step we merge the finest substructure hierarchically, into large, connected regions above the adopted density threshold. This merging process naturally results in a tree-like representation of substructure similar to the dendrograms presented by \cite{rosolowsky2008structural}. 

The main focus of this article is on the parallel implementation of the algorithm which we have described in detail. Our implementation is truly parallel, meaning that it produces \emph{exactly} the same results for varying numbers of MPI tasks. To test the parallelization of \textsc{phew}, we have performed a scaling experiment on a snapshot from a cosmological dark matter simulation. We have found excellent scaling in the relevant range of MPI tasks. When using the same number of MPI tasks that was used for the actual simulation, the runtime of \textsc{phew} $\sim 10 \%$ the time it takes to advance the simulation by one time step. This allows for frequent usage of \textsc{phew} on-the-fly and thus more fine-grained information about how matter assembles in simulations.

\textsc{ramses} has recently been demonstrated to scale well up to 38016 MPI task \citep{alimi2012first} when used to simulate a very large cosmological volume. Even the largest haloes that \textsc{phew} will identify in such a simulation cover only a small fraction of the computational volume. This essentially turns such a setup into a weak scaling experiment for \textsc{phew}, where the scalability is determined by the domain decomposition of \textsc{ramses}. Without having applied \textsc{phew} to such a large setup, we therefore expect the algorithm to show similar scaling properties in this range as the \textsc{ramses} code. A more challenging situation for the \textsc{phew} algorithm is posed by high-resolution zoom simulations of one single halo. In such a situation, the parent halo is spread over almost all MPI tasks, leading to MPI communication across the entire computational domain during the merging process and therefore slightly less favourable scaling properties.

\textsc{phew} has similarities with already existing watershed based halo finders, such as \textsc{denmax} \citep{Bertschinger1991}, \textsc{hop} \citep{Eisenstein1998}, \textsc{skid} \citep{Stadel2001}, \textsc{adaptahop} \citep{Aubert2004}, \textsc{grasshopper} (Potter et al., in prep), but these are either not yet parallelized, do not find substructure or work only on particles. On a first sight, it looks like our approaches to define substructure or parallelization cannot be applied to particle-based data structures since we operate on a mesh-defined density field, while the other codes work on the particle distribution directly. However, the only two concepts that we use which are naturally provided by the grid, namely a local density and the notion of a neighbour, can be also defined for other data structures that do not rely on a grid. Once these properties are defined, the algorithm presented in this paper can be applied to particle data in the same way as we apply it to grid data.  

At the current stage, our implementation of \textsc{phew} is a topological tool only, meaning that it identifies regions in space disregarding physical properties such as the kinetic or gravitational energy of the matter in that volume. For the application of \textsc{phew} as a genuine halo finder, we need to develop an unbinding procedure, which removes dark matter particles from regions they are not gravitationally bound to. We will exploit our hierarchical decomposition into substructure, to pass unbound particle to larger and larger regions, until the particles remain bound. This will unambiguously define the parent halo (or sub-halo) of the particles.

%\the\textwidth
%\the\columnwidth
%\the\textheight

%normal: \fontname\font\ at \the\fontdimen6\font
%\small{small: \the\fontdimen6\font} \\
%\fontname\font at \the\fontdimen6\font

\section*{Acknowledgements}
We want to thank Stephane Colombi for his advice on substructure merging. Furthermore we thank Doug Potter for helpful discussions about programming techniques. The computations leading to this publication have been performed at on the zBox4 and Schroedinger Supercomputers at the University of Zurich and at the Swiss Supercomputing Centre CSCS in Lugano. This work has been supported by the Swiss National Science Foundation SNF under the project ``Computational Astrophysics'' and the PASC co-design project ``Particles and Fields''.

\section*{Competing interests}
The authors declare that they have no competing interests

\section*{Authors' contributions}
AB and RT are the main developers of \textsc{phew} and authors of the manuscript. SC has contributed to the application of \textsc{phew} to particle data in the course of his master thesis. DM was involved in the early development of the algorithm.

\bibliography{clumps}

\begin{thebibliography}{33}
\expandafter\ifx\csname natexlab\endcsname\relax\def\natexlab#1{#1}\fi

\bibitem[{Alimi {et~al}\mbox{.}(2012)Alimi, Bouillot, Rasera, Reverdy,
  Corasaniti, Balmes, Requena, Delaruelle, \& Richet}]{alimi2012first}
Alimi J.-M. {et~al.}, 2012, in Proceedings of the International Conference on
  High Performance Computing, Networking, Storage and Analysis, IEEE Computer
  Society Press, p.~73

\bibitem[{{Arag{\'o}n-Calvo} {et~al}\mbox{.}(2010){Arag{\'o}n-Calvo}, {Platen},
  {van de Weygaert}, \& {Szalay}}]{aragon2010spine}
{Arag{\'o}n-Calvo} M.~A., {Platen} E., {van de Weygaert} R., {Szalay} A.~S.,
  2010, ApJ, 723, 364

\bibitem[{{Aubert} {et~al}\mbox{.}(2004){Aubert}, {Pichon}, \&
  {Colombi}}]{Aubert2004}
{Aubert} D., {Pichon} C., {Colombi} S., 2004, MNRAS, 352, 376

\bibitem[{{Bertschinger} \& {Gelb}(1991)}]{Bertschinger1991}
{Bertschinger} E., {Gelb} J.~M., 1991, Computers in Physics, 5, 164

\bibitem[{Beucher(1994)}]{Beucher1994}
Beucher S., 1994, in Mathematical Morphology And Its Applications To Image
  Processing, Serra J., Soille P., eds.

\bibitem[{Bleuler \& Teyssier(2014)}]{sink_paper}
Bleuler A., Teyssier R., 2014, Monthly Notices of the Royal Astronomical
  Society, 445, 4015

\bibitem[{Davis {et~al}\mbox{.}(1985)Davis, Efstathiou, Frenk, \&
  White}]{davis1985fof}
Davis M., Efstathiou G., Frenk C.~S., White S.~D., 1985, The Astrophysical
  Journal, 292, 371

\bibitem[{{Eisenstein} \& {Hut}(1998)}]{Eisenstein1998}
{Eisenstein} D.~J., {Hut} P., 1998, ApJ, 498, 137

\bibitem[{{Hockney} \& {Eastwood}(1981)}]{Hockney1981}
{Hockney} R.~W., {Eastwood} J.~W., 1981, {Computer Simulation Using Particles}

\bibitem[{{Knebe} {et~al}\mbox{.}(2011){Knebe}, {Knollmann}, {Muldrew},
  {Pearce}, {Aragon-Calvo}, {Ascasibar}, {Behroozi}, {Ceverino}, {Colombi},
  {Diemand}, {Dolag}, {Falck}, {Fasel}, {Gardner}, {Gottl{\"o}ber}, {Hsu},
  {Iannuzzi}, {Klypin}, {Luki{\'c}}, {Maciejewski}, {McBride}, {Neyrinck},
  {Planelles}, {Potter}, {Quilis}, {Rasera}, {Read}, {Ricker}, {Roy},
  {Springel}, {Stadel}, {Stinson}, {Sutter}, {Turchaninov}, {Tweed}, {Yepes},
  \& {Zemp}}]{haloes_mad}
{Knebe} A. {et~al.}, 2011, MNRAS, 415, 2293

\bibitem[{{Knebe} {et~al}\mbox{.}(2013){Knebe}, {Pearce}, {Lux}, {Ascasibar},
  {Behroozi}, {Casado}, {Moran}, {Diemand}, {Dolag}, {Dominguez-Tenreiro},
  {Elahi}, {Falck}, {Gottl{\"o}ber}, {Han}, {Klypin}, {Luki{\'c}},
  {Maciejewski}, {McBride}, {Merch{\'a}n}, {Muldrew}, {Neyrinck}, {Onions},
  {Planelles}, {Potter}, {Quilis}, {Rasera}, {Ricker}, {Roy}, {Ruiz},
  {Sgr{\'o}}, {Springel}, {Stadel}, {Sutter}, {Tweed}, \& {Zemp}}]{knebe_state}
{Knebe} A. {et~al.}, 2013, MNRAS, 435, 1618

\bibitem[{Knollmann \& Knebe(2009)}]{knollmann2009ahf}
Knollmann S.~R., Knebe A., 2009, The Astrophysical Journal Supplement Series,
  182, 608

\bibitem[{Knuth(1998)}]{Knuth}
Knuth D.~E., 1998, The Art of Computer Programming, Vol.~3, Sorting and
  Searching. Addison-Wesley, Reading, Massachusetts

\bibitem[{Meyer(1994)}]{meyer1994topographic}
Meyer F., 1994, Signal processing, 38, 113

\bibitem[{Moga(1997)}]{moga1997parallel}
Moga A., 1997, Parallel watershed algorithms for image segmentation. Tampere
  University of Technology

\bibitem[{Moga \& Gabbouj(1998)}]{moga1998marker}
Moga A.~N., Gabbouj M., 1998, Journal of Parallel and Distributed Computing,
  51, 27

\bibitem[{{Onions} {et~al}\mbox{.}(2013){Onions}, {Ascasibar}, {Behroozi},
  {Casado}, {Elahi}, {Han}, {Knebe}, {Lux}, {Merch{\'a}n}, {Muldrew},
  {Neyrinck}, {Old}, {Pearce}, {Potter}, {Ruiz}, {Sgr{\'o}}, {Tweed}, \&
  {Yue}}]{haloes_notts1}
{Onions} J. {et~al.}, 2013, MNRAS, 429, 2739

\bibitem[{Peng \& Zhang(2011)}]{peng2011automatic}
Peng B., Zhang D., 2011, Image Processing, IEEE Transactions on, 20, 3592

\bibitem[{{Platen} {et~al}\mbox{.}(2007){Platen}, {van de Weygaert}, \&
  {Jones}}]{platen2007cosmic}
{Platen} E., {van de Weygaert} R., {Jones} B.~J.~T., 2007, MNRAS, 380, 551

\bibitem[{{Press} \& {Schechter}(1974)}]{Press1974}
{Press} W.~H., {Schechter} P., 1974, ApJ, 187, 425

\bibitem[{Press {et~al}\mbox{.}(2007)Press, Teukolsky, Vetterling, \&
  Flannery}]{NR}
Press W.~H., Teukolsky S.~A., Vetterling W.~T., Flannery B.~P., 2007, Numerical
  Recipes 3rd Edition: The Art of Scientific Computing, 3rd edn. Cambridge
  University Press, New York, NY, USA

\bibitem[{{Pujol} {et~al}\mbox{.}(2014){Pujol}, {Gazta{\~n}aga}, {Giocoli},
  {Knebe}, {Pearce}, {Skibba}, {Ascasibar}, {Behroozi}, {Elahi}, {Han}, {Lux},
  {Muldrew}, {Neyrinck}, {Onions}, {Potter}, \& {Tweed}}]{haloes_notts2}
{Pujol} A. {et~al.}, 2014, MNRAS, 438, 3205

\bibitem[{Roerdink \& Meijster(2000)}]{Roerdink2000}
Roerdink J.~B., Meijster A., 2000, Fundam. Inf., 41, 187

\bibitem[{Rosolowsky {et~al}\mbox{.}(2008)Rosolowsky, Pineda, Kauffmann, \&
  Goodman}]{rosolowsky2008structural}
Rosolowsky E., Pineda J., Kauffmann J., Goodman A., 2008, The Astrophysical
  Journal, 679, 1338

\bibitem[{Skory {et~al}\mbox{.}(2010)Skory, Turk, Norman, \&
  Coil}]{skory2010parallel}
Skory S., Turk M.~J., Norman M.~L., Coil A.~L., 2010, The Astrophysical Journal
  Supplement Series, 191, 43

\bibitem[{{Springel} {et~al}\mbox{.}(2001){Springel}, {White}, {Tormen}, \&
  {Kauffmann}}]{Springel2001}
{Springel} V., {White} S.~D.~M., {Tormen} G., {Kauffmann} G., 2001, MNRAS, 328,
  726

\bibitem[{{Stadel}(2001)}]{Stadel2001}
{Stadel} J.~G., 2001, PhD thesis, UNIVERSITY OF WASHINGTON

\bibitem[{{Stutzki} \& {Guesten}(1990)}]{Gaussclump}
{Stutzki} J., {Guesten} R., 1990, ApJ, 356, 513

\bibitem[{Sutter {et~al}\mbox{.}(2015)Sutter, Lavaux, Hamaus, Pisani, Wandelt,
  Warren, Villaescusa-Navarro, Zivick, Mao, \& Thompson}]{sutter2015vide}
Sutter P. {et~al.}, 2015, Astronomy and Computing, 9, 1

\bibitem[{{Teyssier}(2002)}]{Teyssier2002}
{Teyssier} R., 2002, AAP, 385, 337

\bibitem[{Way {et~al}\mbox{.}(2014)Way, Gazis, \& Scargle}]{way2014structure}
Way M., Gazis P., Scargle J.~D., 2014, arXiv preprint arXiv:1406.6111

\bibitem[{{Way} {et~al}\mbox{.}(2011){Way}, {Gazis}, \&
  {Scargle}}]{way2011structure}
{Way} M.~J., {Gazis} P.~R., {Scargle} J.~D., 2011, ApJ, 727, 48

\bibitem[{{Williams} {et~al}\mbox{.}(1994){Williams}, {de Geus}, \&
  {Blitz}}]{Clumpfind}
{Williams} J.~P., {de Geus} E.~J., {Blitz} L., 1994, ApJ, 428, 693

\end{thebibliography}

\section*{Appendix A: Glossary}
\paragraph*{Clump: }We use the word clump for the structure after the noise removal. It is the smallest structure that is not considered noise.   
\paragraph*{Key saddle:} The highest saddle point connecting a peak to any neighbouring peak is considered the key saddle. Note that this definition slightly deviates from the one traditionally used in topography.
\paragraph*{Key neighbour:} A peaks key neighbour is the peak it is connected to through the key saddle.
\paragraph*{Neighbouring cell:} Every cell with a common face, edge or corner is considered a neighbour to a given cell.
\paragraph*{Neighbouring peak:} If a cell inside peak patch i is neighbouring a cell in peak patch j, their peaks are considered neighbouring peaks.
\paragraph*{Noise:} A peak with a small relevance (usually less than 1.5) is considered noise.
\paragraph*{Owner:} We denote the MPI task where a given peak is active as the owner of that peak.  
\paragraph*{Peak:}We denote every cell hosting a local density maximum as a peak.
\paragraph*{Peak patch:} Every cell is unambiguously connected to one single density peak by recursively assigning it to the densest neighbouring cell. All cells belonging to a certain peak form the so-called peak patch. The peak patch is the equivalent to the watershed catchment basins for the negative density field.
\paragraph*{Relevance:} The relevance is defined as the ratio of a peaks density to its key saddle density or the density threshold in case of an isolated peak patch. This term is closely related to the topographical term ``prominence'', which denotes the altitude difference of a peak to its highest saddle which connects the peak to a higher neighbour.
\paragraph*{Saddle point:} The density maximum on the connecting surface between two peak patches is located at the saddle point connecting the two peaks.
\paragraph*{Test cell:} Cells with a density above the adopted density threshold are called test cells. Only those are considered in our analysis.

\section*{Appendix B: Algorithmic blocks in pseudocode}

\begin{algorithm*}[b]
{\small
\DontPrintSemicolon
\SetKwFunction{ScatterCommunicate}{ScatterCommunicate}\SetKwFunction{AverageDensity}{AverageDensity}
\SetKwFunction{MPIsum}{MPIsum}\SetKwFunction{GetLocalPeakIndex}{GetLocalPeakIndex}\SetKwFunction{CommunicateSaddlepoints}{CommunicateSaddlepoints}
\SetKwData{alive}{alive}\SetKwData{NewPeak}{NewPeak}\SetKwData{GlobalPeakID}{GlobalPeakID}\SetKwData{KeyNeighbor}{KeyNeighbor}
\SetKwData{KeySaddle}{KeySaddle}\SetKwData{PeakFlag}{PeakPatch}
\SetKwData{SaddleMatrix}{SaddleMatrix}\SetKwData{NewPeak}{NewPeak}\SetKwData{GlobalPeakID}{GlobalPeakID}
\SetKwData{PeakDensity}{PeakDensity}

\For { \rm testcell $\in$ \{testcells\} }{
	\For {\rm neighbour $\in$ \{neighbours\}}{
		\If {\rm(\PeakFlag[neighbour] $\ne$\PeakFlag[testcell]) \textbf{and} (\PeakFlag[neighbour] $>0$)}{
			i$=$\GetLocalPeakIndex(\PeakFlag[testcell])\\
			j$=$\GetLocalPeakIndex(\PeakFlag[neighbour])\\
			\If {\rm \AverageDensity(testcell,neighbour) $>$ \SaddleMatrix[i,j]}{	
				\SaddleMatrix[i,j]=\AverageDensity(testcell,neighbour) \\
				\SaddleMatrix[j,i]=\AverageDensity(testcell,neighbour) 
			}
		}
	}
}
}
\caption{Pseudocode describing the construction of the local saddle point matrices.}\label{algo: saddle point search}
\end{algorithm*}

\begin{algorithm*}
{\small
\DontPrintSemicolon

\SetKwFunction{ScatterCommunicate}{ScatterCommunicate}\SetKwFunction{BuildPeakCommunicator}{BuildPeakCommunicator}
\SetKwFunction{MPIsum}{MPIsum}\SetKwFunction{GetLocalPeakIndex}{GetLocalPeakIndex}\SetKwFunction{CommunicateSaddlepoints}{CommunicateSaddlepoints}
\SetKwData{alive}{alive}\SetKwData{NewPeak}{FinalPeak}\SetKwData{GlobalPeakID}{GlobalPeakID}\SetKwData{KeyNeighbor}{KeyNeighbor}
\SetKwData{KeySaddle}{KeySaddle}
\SetKwData{SaddleMatrix}{SaddleMatrix}\SetKwData{NewPeak}{FinalPeak}\SetKwData{GlobalPeakID}{GlobalPeakID}
\SetKwData{PeakDensity}{PeakDensity}

{\it Preparatory step - initialize two peak-based properties.}\;
\For {\rm  peak $\in$ \{active peaks\}}{
	\alive[peak]=1\;	
	\NewPeak[peak]=\GlobalPeakID[peak]\;	
}
\vspace{\myvskip}

{\it Loop over Levels in the saddle point hierarchy.}\;
mergers=1\;
\While {\rm  mergers $> 0$}{	
mergers=0\;
\vspace{\myvskip}

{\it Propagate the final peak label through key saddle points.}\;
LevelMergers=1\;
\While {\rm LevelMergers $> 0$}{	
	LevelMergers=0\;	
	\CommunicateSaddlepoints\;	
	\BuildPeakCommunicator\;
	\ScatterCommunicate(\PeakDensity,\NewPeak,\alive)\;
	\For {\rm peak $\in$ \{sorted active peaks\}}{		
		\If{\rm \alive[peak]$>0$}{
			PSratio=\PeakDensity[peak]/\KeySaddle[peak]\;
			\If{\rm PSratio $>1.5$ {\bf and} \PeakDensity[\KeyNeighbor[peak] ]$>$ \PeakDensity[peak]}{		
				\NewPeak[peak]=\NewPeak[\KeyNeighbor[peak]]\;
				LevelMergers=LevelMergers+1
			}
		}
	}
	\ScatterCommunicate(\NewPeak)\;
	LevelMergers=\MPIsum(LevelMergers)
}
\vspace{\myvskip}

{\it For every merger, merge the corresponding lines in the saddle point array.}\;
\For {\rm  peak $\in$ \{all peaks\}}{
	\If{\rm \GlobalPeakID[peak] $\ne$ \NewPeak[peak]}{	
		NewIndex$=$\GetLocalPeakIndex(\NewPeak[peak])\;
		\For {\rm column $\in$ \{matrix columns\}}{
			\If{\rm  \SaddleMatrix[peak,column] $>$ \SaddleMatrix[NewIndex,column]}{	
				\SaddleMatrix[NewIndex,column]=\SaddleMatrix[peak,column]\;
				\SaddleMatrix[column,NewIndex]=\SaddleMatrix[peak,column]\;
				}	
			}
		\SaddleMatrix[NewIndex,peak]=0\;	
		\SaddleMatrix[NewIndex,NewIndex]=0\;
		}
	}

\BuildPeakCommunicator\;

\vspace{\myvskip}
{\it Set alive to zero for dead peaks and count mergers.}\;
\For {\rm peak $\in$ \{active peaks\}}{
	\If{\rm \GlobalPeakID[peak] $\ne$ \NewPeak[peak] {\bf and} \alive[peak]==1}{	
		\alive[peak]=0\;
		mergers=mergers+1
	}
}
\ScatterCommunicate(\alive)\;
mergers=\MPIsum(mergers)\;
\vspace{\myvskip}

{\it Remove saddle points linking to dead peaks.}\;
\For {\rm peak $\in$ \{all peaks\}}{
	\For {\rm column $\in$ \{matrix columns\}}{
		\If{\rm  \alive[peak]==0 {\bf or} \alive[column]==0}{
			\SaddleMatrix[peak,column]=0\;
		}
	}
}}}

\caption{Pseudocode describing the parallel merger procedure.}\label{mergeralgo}
\end{algorithm*}

\end{document}